\documentclass[iop]{emulateapj}

\begin{document}

\def\kms{km s$^{-1}$}
\def\msun{M$_{\odot}$}

\slugcomment{2010, ApJ Letters, 719, L23}

\title{ A GALACTIC CENTER ORIGIN FOR HE~0437$-$5439, THE HYPERVELOCITY STAR NEAR THE 
LARGE MAGELLANIC CLOUD\altaffilmark{1}}

\author{Warren R.\ Brown$^1$, Jay Anderson$^2$, Oleg Y.\ Gnedin$^3$, Howard E.\ Bond$^2$,
	Margaret J.\ Geller$^1$, Scott J.\ Kenyon$^1$, and Mario Livio$^2$}

\affil{ $^1$Smithsonian Astrophysical Observatory, 60 Garden St, Cambridge, MA 02138\\
	$^2$Space Telescope Science Institute, 3700 San Martin Dr., Baltimore, MD 21218\\
	$^3$Department of Astronomy, University of Michigan, Ann Arbor, MI 48109 } 

\email{wbrown@cfa.harvard.edu, jayander@stsci.edu, ognedin@umich.edu}

\altaffiltext{1}{Based on observations with the NASA/ESA {\it Hubble Space 
Telescope} obtained at the Space Telescope Science Institute, which is operated by 
the Association of Universities for Research in Astronomy, Inc., under NASA contract 
NAS5-26555.}

\shorttitle{ A Galactic Center Origin for HE~0437$-$5439 }
\shortauthors{Brown et al.}

\begin{abstract}
	We use {\it Hubble Space Telescope} imaging to measure the absolute proper
motion of the hypervelocity star (HVS) HE~0437$-$5439, a short-lived B~star located
in the direction of the Large Magellanic Cloud (LMC). We observe $(\mu_{\alpha},
\mu_{\delta})=(+0.53 \pm0.25({\rm stat})\pm0.33({\rm sys}), +0.09 \pm0.21({\rm
stat})\pm0.48({\rm sys}))$ mas yr$^{-1}$.
	The velocity vector points directly away from the center of the Milky Way; 
an origin from the center of the LMC is ruled out at the 3-$\sigma$ level.  The 
flight time of the HVS from the Milky Way exceeds its main-sequence lifetime, thus 
its stellar nature requires it to be a blue straggler.  The large space velocity 
rules out a Galactic-disk ejection.  Combining the HVS's observed trajectory, 
stellar nature, and required initial velocity, we conclude that HE~0437$-$5439 was 
most likely a compact binary ejected by the Milky Way's central black hole.

\end{abstract}


\section{INTRODUCTION}

	Hypervelocity stars (HVSs) are stars escaping the Milky Way.  Predicted by 
\citet{hills88} as a natural consequence of the massive black hole (MBH) in the 
Galactic center, HVSs were first discovered by \citet{brown05}.  The proposed 
ejection mechanism is a three-body interaction between binary stars and a MBH, or 
possibly a pair of MBHs \citep{yu03}, that ejects HVSs from the Milky Way at a rate 
of 10$^{-3}$ to 10$^{-4}$ yr$^{-1}$ \citep{perets07}.  Approximately 5\% of HVS 
ejections should be compact binaries \citep{lu07, perets09a}.
	At least 14 unbound stars have now been discovered in a targeted survey for
HVSs \citep{brown06, brown06b, brown07a, brown07b, brown09a} and another 3-5 unbound
stars discovered in other surveys \citep{edelmann05, hirsch05, heber08, tillich09,
irrgang10}.  Unlike high proper-motion pulsars, which are the remnants of supernova
explosions, known HVSs are mostly B-type main-sequence stars \citep{fuentes06,
bonanos08, lopezmorales08, przybilla08, przybilla08b}.  Similar B stars are observed
in short-period orbits around the central MBH \citep{ghez08, gillessen09} and may be
the former companions of HVSs \citep{gould03b, ginsburg06}.  But, until now, the
evidence linking HVSs to the Galactic center has been indirect.

	One of the most intriguing HVSs is HE 0437$-$5439 \citep{edelmann05}, called 
HVS3 in the catalog of \citet{brown09a}.  HVS3 is a 9 \msun\ B star located 
16$\arcdeg$ from LMC on the sky.  It has a heliocentric radial velocity of +723 
\kms\ and a heliocentric distance of 61 kpc.  The flight time of HVS3 from the Milky 
Way is about 100 Myr, substantially more than its $\simeq$20 Myr main-sequence 
lifetime. \citet{edelmann05} consider the possibility that HVS3 was ejected from the 
Milky Way as a binary and evolved into a 9 \msun\ blue straggler, but conclude that 
an ejection from the LMC is more likely.

	An LMC origin requires that HVS3 was ejected at nearly 1000 \kms\
\citep{przybilla08}, a velocity comparable to the $\sim$1000 \kms\ escape velocity
from the surface of the 9 \msun\ star.  \citet{gualandris07} suggest that an
intermediate-mass black hole (IMBH) in a massive young LMC star cluster can provide
the required ejection velocity without disrupting the star. Alternatively,
\citet{gvaramadze08, gvaramadze09} suggest that HVS3 was ejected by a stellar
binary-binary encounter involving $>$100 \msun\ stars.  The full space motion of
HVS3 constrains its origin.

	We use the {\it Hubble Space Telescope} ({\it HST}) to measure the absolute 
proper motion of HVS3, the first result from an astrometric campaign to measure 
proper motions for all HVSs.  The velocity vector of HVS3 points directly from the 
Galactic center.  Summing the random and systematic proper motion errors, an origin 
from the center of the LMC is ruled out at the 3-$\sigma$ level.  We bring to bear 
three observations to restrict the origin of HVS3: its trajectory, stellar nature, 
and required initial velocity.  We rule out a Galactic-disk ejection, and conclude 
that HVS3 was once a binary ejected by the Milky Way's central MBH as proposed by 
\citet{perets09a}.

\begin{figure*} 
 \centerline{\includegraphics[width=4.8in]{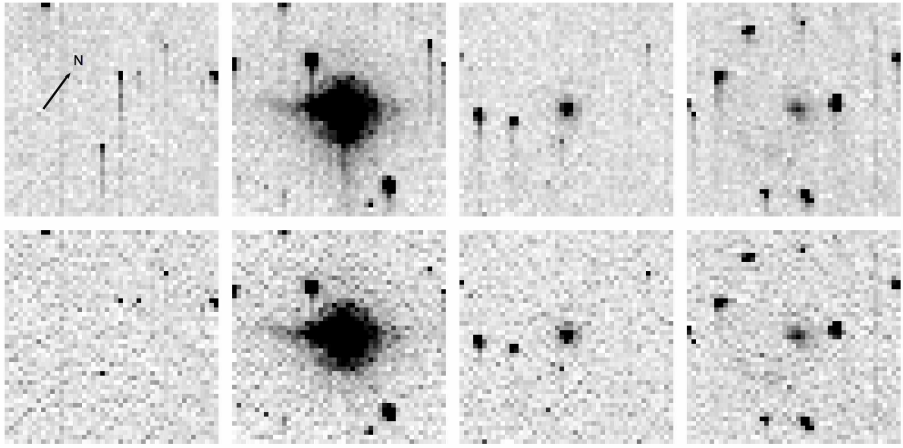}}
 \caption{ \label{fig:cti}
	Top row:  Uncorrected thumbnail images from \hbox{epoch-2} showing, from
left to right, warm pixels, a bright star, a faint star, and a faint galaxy.  CTI 
trails in the (vertical) readout direction are present in all objects.
	Bottom row:  Corrected images, showing CTI trails restored to their rightful 
pixels.  Readnoise is slightly amplified because the CTI correction is a mild 
deconvolution, but this adds very little random error to the astrometry.}
 \end{figure*}

\section{PROPER MOTION}

	We measure the proper motion of HVS3 using two epochs of {\it HST} Advanced
Camera for Surveys (ACS) imaging obtained on 2006 Jul 8 and 2009 Dec 23.  Each epoch
consists of six dithered $\simeq$5 min exposures in the F850LP filter, designed to
maximize the signal of the background galaxies and to prevent the bright HVS from
saturating.

	Charge-transfer inefficiency (CTI) is the largest source of uncertainty in
our proper motion measurement.  Physically, CTI is due to pixel defects temporarily
trapping electrons as they are shifted across the detector during readout.  The
consequence is image distortion (seen in Figure \ref{fig:cti}):  up to 20\% of a
background galaxy's electrons can be trailed across neighboring pixels in the
readout direction \citep{anderson10}.  We can minimize the effects of CTI by
obtaining images at the same telescope orientation, but unfortunately our
\hbox{epoch-2} images had to be rotated by 180$\arcdeg$ relative to our
\hbox{epoch-1} images because the \hbox{epoch-1} guidestars were no longer
available.  This mis-orientation magnifies the CTI effect and requires that we
correct for it.

	We use a pixel-based CTI correction recently developed by 
\citet{anderson10}.  Application of this technique to a 47 Tuc field demonstrates 
that it mitigates CTI errors to better than 75\% in terms of photometry, astrometry, 
and shape.  Figure \ref{fig:cti} illustrates the results of CTI correction in one of 
our \hbox{epoch-2} images.  We apply 55\% of the \hbox{epoch-2} CTI correction to 
the \hbox{epoch-1} images, because ACS had been in space about half as long.  A 
study of the warm pixels in dark frames contemporaneous to \hbox{epoch-1} validates 
this strategy.  If we perform no correction for known CTI effects, we would measure 
the wrong location for HVS3:  its displacement would shift 1.8 mas yr$^{-1}$ north 
and 1.2 mas yr$^{-1}$ east.

	We begin our astrometric measurement by stacking the \hbox{epoch-2} images,
using bright stars to transform the six exposures into a common distortion-corrected
frame \citep{anderson06}.  We supersample this frame 2 times relative to the ACS
pixel scale.  We select 17 stars and 18 compact galaxies and use their stacked
images as templates.  The templates enable us to measure a consistent position for
each object in each exposure \citep[see][]{mahmud08}.  We then define an absolute
frame with north up, east on the left, and a scale of 50 mas pix$^{-1}$.  We use the
pipeline {\tt \_drz} images from \hbox{epoch-1} to measure rough positions for each
galaxy and, using the galaxies alone, we determine the transformation from the
distortion-corrected frame (Anderson \& King ISR/ACS 04-15) of each exposure into
this absolute reference frame.  We refine the reference frame and the
transformations by taking the average transformed galaxy positions as the new galaxy
positions and iterating.  A few iterations yield a set of 11 high-quality galaxies
with RMS position errors $<$12 mas in the reference frame and $<$18 mas residuals in
each individual frame where they are well measured.

\begin{figure} 
 \includegraphics[width=3.3in]{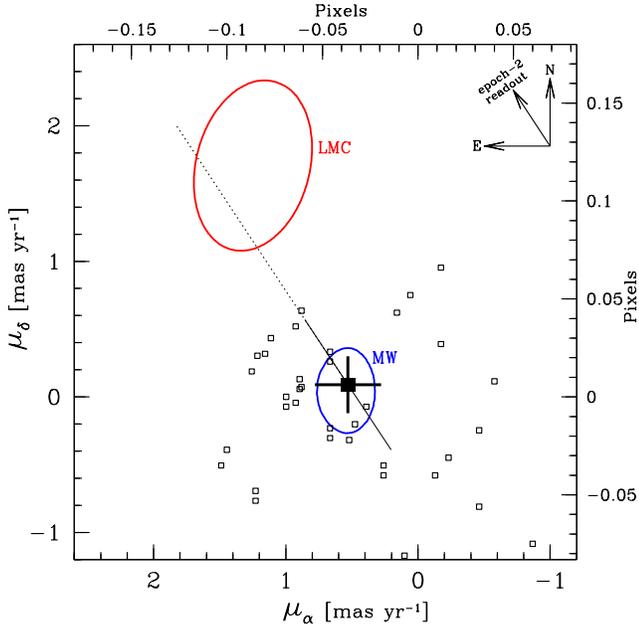}
 \caption{ \label{fig:mu}
	Mean proper motion of HVS3 ({\it solid square}) and the distribution of 
proper motions ({\it open squares}) measured from the individual epoch-1 and epoch-2 
images.  The 1-$\sigma$ statistical uncertainty ({\it plus sign}) and systematic 
uncertainty ({\it solid line}) are indicated, as well as the full CTI correction 
({\it dotted line}). We also show the locus of proper motions with trajectories 
passing within 8 kpc of the Milky Way center ({\it lower ellipse}) and within 3 kpc 
of the LMC center ({\it upper ellipse}) at the time of pericenter passage.  HVS3's 
velocity vector points from the Milky Way; an origin from the center of the LMC is 
ruled out at 3-$\sigma$ (systematic). }
 \end{figure}

	We base our final astrometric transformations exclusively on the positions 
of the 11 high-quality galaxies; thus inter-epoch displacements in our reference 
frame correspond directly to HVS3's absolute proper motion.  Figure \ref{fig:mu} 
plots the 36 displacements found from comparing each of the six \hbox{epoch-1} 
images against each of the six \hbox{epoch-2} images.  The average displacement is 
0.0375$\pm$0.017 pix over the 3.46 yr baseline, which corresponds to 
$(\mu_{\alpha},\mu_{\delta}) = (+0.53\pm0.25, +0.09\pm0.21)$ mas yr$^{-1}$.  Here we 
quote the 1-$\sigma$ error-in-the-mean, calculated under the conservative assumption 
that there are 6 independent \hbox{epoch-1} -- \hbox{epoch2} differences.  As a 
check, we compute HVS3's displacement using two bright galaxies directly east and 
west of the object and find an answer consistent at the 1-$\sigma$ level. The 
statistical error-in-the-mean is an underestimate of the total uncertainty, however.  
The CTI correction may be off by as much as 25\%; we conservatively estimate a 
1-$\sigma$ systematic error of $\pm$0.58 mas yr$^{-1}$ in the readout direction, 
oriented $34.17\arcdeg$ east of north.

	We transform HVS3's observed heliocentric motion into the Galactic rest 
frame assuming a circular velocity of 220 \kms\ and a solar motion of 
$(U,V,W)=(11.1,12.24,7.25)$ \kms\ per \citet{schonrich10}.  The Galactocentric 
velocity components are thus $(U,V,W)=(-119, -400, -355)$ \kms .

\section{TRAJECTORY}

	We calculate the trajectory of HVS3 with respect to the Milky Way and LMC 
using the \citet{kenyon08} three-component potential model.  This model uses a 
circular velocity of 220 \kms\ at the solar radius and assumes a total stellar mass 
of $4.4\times10^{10}$ \msun\ and a dark matter halo virial mass of $1\times10^{12}$ 
\msun\ with a scale radius of 20 kpc. The details of the mass distribution are not 
very important because HVS3 spends most its time traveling through the outer halo; a 
cross-check using the \citet{gnedin05} potential yields identical results.

	We account for the changing position of the LMC in all of our calculations.  
We assume the LMC currently has the location and line-of-sight velocity from
\citet{vandermarel01} and the proper motion from \citet{kallivayalil06}.  We
establish the LMC's location versus time by computing the LMC's orbit in our
potential model.  We then assign the LMC a point mass of $2\times10^{10}$ \msun\ in
our calculations.

	To put the proper motion of HVS3 in context, we begin by computing 
trajectories for the full range of possible proper motions.  The elliptical contours 
in Figure \ref{fig:mu} show the locus of proper motions whose outbound trajectories 
pass within 8 kpc of the Milky Way center (lower ellipse) and within 3 kpc of the 
LMC center (upper ellipse) at the time of pericenter passage.  We chose 3 kpc for 
the LMC because this radius encompasses the extent of the LMC bar, 85\% of the LMC's 
observed young stellar objects \citep{gruendl09}, and all of the young clusters 
proposed by \citet{gualandris07} for the origin of HVS3.

\begin{deluxetable}{llr}        
\tablewidth{0pt}
\tablecaption{HVS3 OBSERVED PARAMETERS}
\tablecolumns{3}
\tablehead{ \colhead{Parameter} & \colhead{Value} & \colhead{Ref} }
        \startdata
RA (J2000)              & 4:38:12.8             & 1 \\
Dec (J2000)             & $-$54:33:12           & 1 \\
$v_{\rm helio}$ (\kms ) & 723 $\pm$ 3           & 1,2,3 \\
$V$ (mag)               & 16.36 $\pm$ 0.04      & 2 \\
$B-V$ (mag)             & $-$0.23 $\pm$ 0.03    & 2 \\
$T_{\rm eff}$ (K)       & 23000 $\pm$ 1000      & 3 \\
$\log{g}$ (cgs)         &  3.95 $\pm$ 0.10      & 3 \\
$v\sin{i}$ (\kms )      & 55 $\pm$ 2            & 2,3 \\
$M$ (\msun )            & 9.1 $\pm$ 0.8         & 3 \\
$d_{\rm helio}$ (kpc)   & 61 $\pm$ 9            & 3 \\
$\mu_{\alpha}$ (mas yr$^{-1}$) & +0.53 $\pm$0.25(stat)$\pm$0.33(sys) & \\
$\mu_{\delta}$ (mas yr$^{-1}$) & +0.09 $\pm$0.21(stat)$\pm$0.48(sys) &
        \enddata
\tablerefs{ (1) \citet{edelmann05}; (2) \citet{bonanos08}; (3) \citet{przybilla08} }
\tablecomments{All errors are 1-$\sigma$}
 \end{deluxetable}

	Figure \ref{fig:mu} shows that the full space motion of HVS3 points directly 
from the Milky Way.  The Galactic center and most of the disk fall within the proper 
motion 1-$\sigma$ systematic error bar.  An LMC origin, on the other hand, requires 
a large, northerly tangential motion ruled out at 3-$\sigma$ (systematic) or better.

\begin{figure*} 
 \centerline{\includegraphics[width=4.0in]{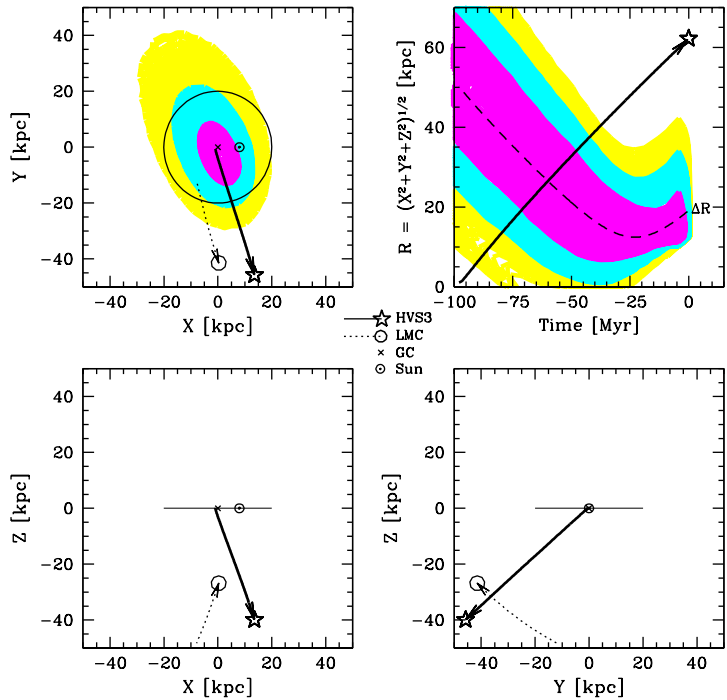}}
 \caption{ \label{fig:orbit}
	Trajectory of HVS3 ({\it solid line}) and LMC ({\it dotted line}) over the
past 98 Myr plotted in physical Cartesian coordinates centered on the Milky Way.  
Symbols mark the present position of HVS3 ({\it star}), the LMC ({\it open circle}),
and the Sun ($\odot$).  Arrowheads indicate the present direction of motion.  The
Milky Way disk is indicated by a 20 kpc radius circle; the LMC is indicated by a 3
kpc radius circle.  The likelihood of HVS3's Galactic plane-crossing is plotted for
1-$\sigma$ ({\it magenta ellipse}), 2-$\sigma$ ({\it cyan ellipse}), and 3-$\sigma$
({\it yellow ellipse}) levels in the upper left panel.  The separation of HVS3 and
the center of the LMC ({\it dashed line}) and the 1-$\sigma$ ({\it magenta band}),
2-$\sigma$ ({\it cyan band}), and 3-$\sigma$ ({\it yellow band}) distribution of
separations are plotted in the upper right panel.  All errors, statistical and
systematic, are considered simultaneously. }
 \end{figure*}

	Next we consider the trajectory of HVS3 in physical space.  Figure
\ref{fig:orbit} plots the present location and computed trajectory for both HVS3 and
the LMC in a Cartesian coordinate system centered on the Milky Way.  Arrowheads
indicate the present direction of motion.
	23 Myr ago, HVS3 passed within 13 kpc of the LMC.  For comparison, the LMC's 
outermost stars and star clusters are 10 kpc distant from the LMC 
\citep{vandermarel01, bica96}.  98 Myr ago, HVS3 crossed the Galactic plane.
	\citet{mcmillan10} argue that the Sun's circular velocity should be 240 
\kms\ at R$_0$=8 kpc; this changes the flight time and Galactic plane-crossing 
location of HVS3 by $<$3 Myr and $<$1 kpc, respectively.

	To establish where HVS3 most likely crossed the Galactic plane, we propagate 
the measurement uncertainties (Table 1) through a Monte Carlo calculation.  We 
assume that the uncertainties are Gaussian distributed, and calculate trajectories 
for 1 million random draws of HVS3's distance, radial velocity, and proper motion.  
Figure \ref{fig:orbit} shows the resulting distribution of Galactic plane-crossing 
locations:  the ellipses show the 1-, 2-, and 3-$\sigma$ levels of the distribution, 
considering all the errors simultaneously.  The $\pm$9 kpc distance uncertainty is a 
dominant source of error.  The origin of HVS3 is formally consistent with both the 
Galactic center and the Galactic disk based on its trajectory alone.

\section{STELLAR NATURE}

	\citet{bonanos08} and \citet{przybilla08} independently measure solar Fe and
half-solar C, N, O, Mg, Si abundances for HVS3.  The observed abundance pattern is
consistent with an LMC or a Milky Way outer disk origin, however the pattern is also
consistent with a Galactic center origin within the 1- to 2-$\sigma$ uncertainties
\citep{przybilla08}.  If HVS3 is a blue straggler, on the other hand, it is unclear
what abundance pattern we should expect.  Fortunately, the kinematics paint a
clearer picture.

	\citet{bonanos08} and \citet{przybilla08} find that the surface gravity, 
effective temperature, and rotation of HVS3 are those of a 9 \msun\ main-sequence 
star with a lifetime of $\simeq$20 Myr.  This lifetime, combined with its 98$\pm$17 
Myr flight time from the Milky Way, requires that HVS3 is a blue straggler.

	Blue stragglers are ubiquitous in globular clusters, open clusters, and in 
the ``field'' of the Milky Way:  half of all A-type stars in the stellar halo are 
main-sequence blue stragglers \citep[e.g][]{wilhelm99b, preston00, brown08b}.  Blue 
stragglers probably form by either binary mass transfer or mergers \citep{bailyn95}.  
The resulting blue straggler then evolves quite normally as a main-sequence star of 
higher mass \citep{sills09}.  For HVS3, the implication is that a pair of 4.5 \msun\ 
stars, with main-sequence lifetimes of 160 Myr \citep{girardi02}, evolved and merged 
during their 98 Myr flight time from the Milky Way.

	The only alternative is that HVS3 is a post main-sequence star 
\citep{demarque07}.  However, \citet{edelmann05} show that HVS3's $\log{g}$ is too 
low for its temperature, and its rotation too large, for it to be a horizontal 
branch (HB) star.  Alternatively, HVS3 could be an AGB-manqu\'e star, a post-HB star 
with too little hydrogen to reach the asymptotic giant branch (AGB), or a post-AGB 
star.  Evolutionary tracks show that AGB-manqu\'e and post-AGB stars spend 10$^5$ 
yrs at HVS3's $T_{\rm eff}$ and $\log{g}$, with approximately one-tenth the 
luminosity of a main-sequence star \citep{dorman93}.  Thus in both the AGB-manqu\'e 
and post-AGB scenarios, HVS3's lifetime constraint is relaxed but the star is {\it 
still approaching} the LMC.  The AGB-manqu\'e and post-AGB interpretations are 
problematic, however.  HVS3 rotates too rapidly for such an evolved star.  In 
addition, the joint probability of finding a star in a rare evolutionary phase that 
lasts $<$10$^{-5}$ of the star's lifetime and that is also a HVS ($\sim$10$^{-7}$ of 
Milky Way stars are HVSs) is very small.  We find no compelling reason to contradict 
the observed main-sequence nature of HVS3, and assume that HVS3 is a blue straggler.

\section{ORIGIN}

	Because HVS3 is a blue straggler with a main-sequence lifetime less than its 
flight time, it must have been ejected from the Milky Way as a binary system that 
subsequently merged during its flight.

	It is difficult, however, to eject a binary with the necessary velocity from 
the disk of the Milky Way.
	The trajectory of HVS3 crosses the disk at 820 \kms . In comparison, a pair 
of 4.5 \msun\ stars with 3.4 R$_{\odot}$ radii and separated by 8.5 R$_{\odot}$ 
(their Roche-lobe overflow separation) have an orbital velocity of 320 \kms .  Thus 
a 820 \kms\ kick exceeds by 2.5 times the orbital velocity of a pair of 4.5 \msun\ 
stars in a compact binary, and it also exceeds the 710 \kms\ escape velocity from 
the surface of a 4.5 \msun\ star.
	Thus any stellar dynamical mechanism that might have ejected the progenitor 
binary from the disk would have destroyed the progenitor.

	Ejecting a binary at 820 \kms\ likely requires a more massive and compact 
object, such as a MBH.  Two proposed mechanisms are:  a binary MBH that ejects 
stellar binaries as HVSs \citep{lu07}, and a single MBH that ejects binaries by 
triple disruption \citep{perets09a}.

	Provided that the separation between a pair of MBHs is substantially larger 
than the tidal disruption distance of binary stars, \citet{lu07} estimate a that a 
binary MBH ejects $>$50\% of incoming binaries with $v>900$ \kms .  They predict 
that HVS binaries comprise 3-5\% of all HVSs.  \citet{sesana09} find similar HVS 
binary frequencies based on more detailed simulations, although \citet{perets09c} 
argues this may be an overestimate.  Interestingly, the 16-19 HVSs observed to date 
are consistent with finding 1 HVS binary.
	In the context of HVS3, the binary MBH scenario implies that an IMBH 
in-spiraled into the Milky Way's central MBH about 100 Myr ago.

	\citet{perets09a}, on the other hand, argues that HVS binary ejection is the
natural consequence of a single MBH disrupting triple stars.  For context,
essentially all O and B stars are observed in binaries, and half of the binaries are
equal mass twins \citep{pinsonneault06}.  Hierarchical triples are stable if the
outer binary has a much larger separation than the inner compact binary;  
\citet{perets09a} shows that a MBH interaction can disrupt the outer binary while
the inner binary remains bound.  Perets estimates that 3-5\% of HVS ejections will
be HVS binaries from disrupted triples.  Because the ejected binaries are compact,
most will undergo type-A mass transfer and merge.  Thus the \citet{perets09a} triple
disruption mechanism also explains the blue-straggler nature of HVS3.

	If HVSs originate from the Galactic center, it appears inevitable that some
HVSs will be binaries.  We estimate the likelihood of HVS3 being a (former) binary
by looking at the complete survey of \citet{brown07b}, who infer 96$\pm$20 3-4
\msun\ HVSs are located within 100 kpc of the Galactic center.  If 5\% of HVS
ejections are binaries, then there should be 5 HVS binaries containing $\simeq$4
\msun\ stars.  Assuming these (necessarily compact) binaries all evolve to form a
$\simeq$8 \msun\ blue straggler with a lifetime of 30 Myr, we expect 30 Myr / 160
Myr = 20\% in the blue-straggler phase, or about 1 star on average.  We conclude
that the stellar nature and kinematics of HVS3 are consistent with it being an
ejected HVS binary as proposed by \citet{lu07} and \citet{perets09a}.

\section{CONCLUSION}

	We measure the proper motion of HVS3 and find that its velocity vector
points from the Milky Way.  An origin from the center of the LMC is ruled out at
3-$\sigma$ (systematic).  Because HVS3's travel time from the Milky Way exceeds its
main-sequence lifetime, it must be a blue straggler whose progenitor was ejected
from the Milky Way as a binary system.  The star's abundance pattern is ambiguous,
but the kinematics are clear.  The finite size of stars imposes a velocity
constraint that rules out a disk ejection.  Combining the observed trajectory,
stellar nature, and required initial velocity, we conclude that HVS3 is a former
binary ejected by the MBH in the Galactic center.
	In the future, a 3rd epoch of imaging will greatly improve the proper motion 
constraint:  doubling the time baseline will halve the internal error, and matching 
the image orientation will reduce the systematic error by 67\%.

\acknowledgements

	We thank Andy Gould, Hagai Perets, and the anonymous referee for helpful 
comments.  Support for this research was provided by NASA through grants GO-10824 
and GO-11782 from the Space Telescope Science Institute, which is operated by the 
Association of Universities for Research in Astronomy, Inc., under NASA contract 
NAS5-26555. This research makes use of NASA's Astrophysics Data System Bibliographic 
Services.  This work was supported in part by the Smithsonian Institution.

{\it Facilities:} \facility{HST (ACS)}


\begin{thebibliography}{50}
\expandafter\ifx\csname natexlab\endcsname\relax\def\natexlab#1{#1}\fi

\bibitem[{{Anderson} \& {King}(2006)}]{anderson06}
{Anderson}, J. \& {King}, I.~R. 2006, {``PSFs, Photometry, and Astronomy for
  the ACS/WFC''}, Tech. rep., STScI

\bibitem[{{Anderson} \& {Bedin}(2010)}]{anderson10}
{Anderson}, J.~W. \& {Bedin}, L.~R. 2010, \pasp, accepted

\bibitem[{{Bailyn}(1995)}]{bailyn95}
{Bailyn}, C.~D. 1995, \araa, 33, 133

\bibitem[{{Bica} {et~al.}(1996){Bica}, {Claria}, {Dottori}, {Santos}, \&
  {Piatti}}]{bica96}
{Bica}, E., {Claria}, J.~J., {Dottori}, H., {Santos}, Jr., J.~F.~C., \&
  {Piatti}, A.~E. 1996, \apjs, 102, 57

\bibitem[{{Bonanos} {et~al.}(2008){Bonanos}, {L{\'o}pez-Morales}, {Hunter}, \&
  {Ryans}}]{bonanos08}
{Bonanos}, A.~Z., {L{\'o}pez-Morales}, M., {Hunter}, I., \& {Ryans}, R.~S.~I.
  2008, \apjl, 675, L77

\bibitem[{{Brown} {et~al.}(2008){Brown}, {Beers}, {Wilhelm}, {Allende Prieto},
  {Geller}, {Kenyon}, \& {Kurtz}}]{brown08b}
{Brown}, W.~R., {Beers}, T.~C., {Wilhelm}, R., {Allende Prieto}, C., {Geller},
  M.~J., {Kenyon}, S.~J., \& {Kurtz}, M.~J. 2008, \aj, 135, 564

\bibitem[{{Brown} {et~al.}(2009){Brown}, {Geller}, \& {Kenyon}}]{brown09a}
{Brown}, W.~R., {Geller}, M.~J., \& {Kenyon}, S.~J. 2009, \apj, 690, 1639

\bibitem[{{Brown} {et~al.}(2005){Brown}, {Geller}, {Kenyon}, \&
  {Kurtz}}]{brown05}
{Brown}, W.~R., {Geller}, M.~J., {Kenyon}, S.~J., \& {Kurtz}, M.~J. 2005,
  \apjl, 622, L33

\bibitem[{{Brown} {et~al.}(2006{\natexlab{a}}){Brown}, {Geller}, {Kenyon}, \&
  {Kurtz}}]{brown06}
---. 2006{\natexlab{a}}, \apjl, 640, L35

\bibitem[{{Brown} {et~al.}(2006{\natexlab{b}}){Brown}, {Geller}, {Kenyon}, \&
  {Kurtz}}]{brown06b}
---. 2006{\natexlab{b}}, \apj, 647, 303

\bibitem[{{Brown} {et~al.}(2007{\natexlab{a}}){Brown}, {Geller}, {Kenyon},
  {Kurtz}, \& {Bromley}}]{brown07a}
{Brown}, W.~R., {Geller}, M.~J., {Kenyon}, S.~J., {Kurtz}, M.~J., \& {Bromley},
  B.~C. 2007{\natexlab{a}}, \apj, 660, 311

\bibitem[{{Brown} {et~al.}(2007{\natexlab{b}}){Brown}, {Geller}, {Kenyon},
  {Kurtz}, \& {Bromley}}]{brown07b}
---. 2007{\natexlab{b}}, \apj, 671, 1708

\bibitem[{{Demarque} \& {Virani}(2007)}]{demarque07}
{Demarque}, P. \& {Virani}, S. 2007, \aap, 461, 651

\bibitem[{{Dorman} {et~al.}(1993){Dorman}, {Rood}, \& {O'Connell}}]{dorman93}
{Dorman}, B., {Rood}, R.~T., \& {O'Connell}, R.~W. 1993, \apj, 419, 596

\bibitem[{{Edelmann} {et~al.}(2005){Edelmann}, {Napiwotzki}, {Heber},
  {Christlieb}, \& {Reimers}}]{edelmann05}
{Edelmann}, H., {Napiwotzki}, R., {Heber}, U., {Christlieb}, N., \& {Reimers},
  D. 2005, \apjl, 634, L181

\bibitem[{{Fuentes} {et~al.}(2006){Fuentes}, {Stanek}, {Gaudi}, {McLeod},
  {Bogdanov}, {Hartman}, {Hickox}, \& {Holman}}]{fuentes06}
{Fuentes}, C.~I., {Stanek}, K.~Z., {Gaudi}, B.~S., {McLeod}, B.~A., {Bogdanov},
  S., {Hartman}, J.~D., {Hickox}, R.~C., \& {Holman}, M.~J. 2006, \apjl, 636,
  L37

\bibitem[{{Ghez} {et~al.}(2008)}]{ghez08}
{Ghez}, A.~M. {et~al.} 2008, \apj, 689, 1044

\bibitem[{{Gillessen} {et~al.}(2009){Gillessen}, {Eisenhauer}, {Trippe},
  {Alexander}, {Genzel}, {Martins}, \& {Ott}}]{gillessen09}
{Gillessen}, S., {Eisenhauer}, F., {Trippe}, S., {Alexander}, T., {Genzel}, R.,
  {Martins}, F., \& {Ott}, T. 2009, \apj, 692, 1075

\bibitem[{{Ginsburg} \& {Loeb}(2006)}]{ginsburg06}
{Ginsburg}, I. \& {Loeb}, A. 2006, \mnras, 368, 221

\bibitem[{{Girardi} {et~al.}(2002)}]{girardi02}
{Girardi}, L. {et~al.} 2002, \aap, 391, 195

\bibitem[{{Gnedin} {et~al.}(2005){Gnedin}, {Gould}, {Miralda-Escud{\'e}}, \&
  {Zentner}}]{gnedin05}
{Gnedin}, O.~Y., {Gould}, A., {Miralda-Escud{\'e}}, J., \& {Zentner}, A.~R.
  2005, \apj, 634, 344

\bibitem[{{Gould} \& {Quillen}(2003)}]{gould03b}
{Gould}, A. \& {Quillen}, A.~C. 2003, \apj, 592, 935

\bibitem[{{Gruendl} \& {Chu}(2009)}]{gruendl09}
{Gruendl}, R.~A. \& {Chu}, Y. 2009, \apjs, 184, 172

\bibitem[{{Gualandris} \& {Portegies Zwart}(2007)}]{gualandris07}
{Gualandris}, A. \& {Portegies Zwart}, S. 2007, \mnras, 376, L29

\bibitem[{{Gvaramadze} {et~al.}(2008){Gvaramadze}, {Gualandris}, \& {Portegies
  Zwart}}]{gvaramadze08}
{Gvaramadze}, V.~V., {Gualandris}, A., \& {Portegies Zwart}, S. 2008, \mnras,
  385, 929

\bibitem[{{Gvaramadze} {et~al.}(2009){Gvaramadze}, {Gualandris}, \& {Portegies
  Zwart}}]{gvaramadze09}
---. 2009, \mnras, 396, 570

\bibitem[{{Heber} {et~al.}(2008){Heber}, {Edelmann}, {Napiwotzki}, {Altmann},
  \& {Scholz}}]{heber08}
{Heber}, U., {Edelmann}, H., {Napiwotzki}, R., {Altmann}, M., \& {Scholz},
  R.-D. 2008, \aap, 483, L21

\bibitem[{{Hills}(1988)}]{hills88}
{Hills}, J.~G. 1988, \nat, 331, 687

\bibitem[{{Hirsch} {et~al.}(2005){Hirsch}, {Heber}, {O'Toole}, \&
  {Bresolin}}]{hirsch05}
{Hirsch}, H.~A., {Heber}, U., {O'Toole}, S.~J., \& {Bresolin}, F. 2005, \aap,
  444, L61

\bibitem[{{Irrgang} {et~al.}(2010){Irrgang}, {Przybilla}, {Heber}, {Fernanda
  Nieva}, \& {Schuh}}]{irrgang10}
{Irrgang}, A., {Przybilla}, N., {Heber}, U., {Fernanda Nieva}, M., \& {Schuh},
  S. 2010, \apj, 711, 138

\bibitem[{{Kallivayalil} {et~al.}(2006){Kallivayalil}, {van der Marel},
  {Alcock}, {Axelrod}, {Cook}, {Drake}, \& {Geha}}]{kallivayalil06}
{Kallivayalil}, N., {van der Marel}, R.~P., {Alcock}, C., {Axelrod}, T.,
  {Cook}, K.~H., {Drake}, A.~J., \& {Geha}, M. 2006, \apj, 638, 772

\bibitem[{{Kenyon} {et~al.}(2008){Kenyon}, {Bromley}, {Geller}, \&
  {Brown}}]{kenyon08}
{Kenyon}, S.~J., {Bromley}, B.~C., {Geller}, M.~J., \& {Brown}, W.~R. 2008,
  \apj, 680, 312

\bibitem[{{L{\'o}pez-Morales} \& {Bonanos}(2008)}]{lopezmorales08}
{L{\'o}pez-Morales}, M. \& {Bonanos}, A.~Z. 2008, \apjl, 685, L47

\bibitem[{{Lu} {et~al.}(2007){Lu}, {Yu}, \& {Lin}}]{lu07}
{Lu}, Y., {Yu}, Q., \& {Lin}, D.~N.~C. 2007, \apjl, 666, L89

\bibitem[{{Mahmud} \& {Anderson}(2008)}]{mahmud08}
{Mahmud}, N. \& {Anderson}, J. 2008, \pasp, 120, 907

\bibitem[{{McMillan} \& {Binney}(2010)}]{mcmillan10}
{McMillan}, P.~J. \& {Binney}, J.~J. 2010, \mnras, 402, 934

\bibitem[{{Perets}(2009{\natexlab{a}})}]{perets09c}
{Perets}, H.~B. 2009{\natexlab{a}}, \apj, 690, 795

\bibitem[{{Perets}(2009{\natexlab{b}})}]{perets09a}
---. 2009{\natexlab{b}}, \apj, 698, 1330

\bibitem[{{Perets} {et~al.}(2007){Perets}, {Hopman}, \& {Alexander}}]{perets07}
{Perets}, H.~B., {Hopman}, C., \& {Alexander}, T. 2007, \apj, 656, 709

\bibitem[{{Pinsonneault} \& {Stanek}(2006)}]{pinsonneault06}
{Pinsonneault}, M.~H. \& {Stanek}, K.~Z. 2006, \apjl, 639, L67

\bibitem[{{Preston} \& {Sneden}(2000)}]{preston00}
{Preston}, G.~W. \& {Sneden}, C. 2000, \aj, 120, 1014

\bibitem[{{Przybilla} {et~al.}(2008{\natexlab{a}}){Przybilla}, {Nieva},
  {Heber}, {Firnstein}, {Butler}, {Napiwotzki}, \& {Edelmann}}]{przybilla08}
{Przybilla}, N., {Nieva}, M.~F., {Heber}, U., {Firnstein}, M., {Butler}, K.,
  {Napiwotzki}, R., \& {Edelmann}, H. 2008{\natexlab{a}}, \aap, 480, L37

\bibitem[{{Przybilla} {et~al.}(2008{\natexlab{b}}){Przybilla}, {Nieva},
  {Tillich}, {Heber}, {Butler}, \& {Brown}}]{przybilla08b}
{Przybilla}, N., {Nieva}, M.~F., {Tillich}, A., {Heber}, U., {Butler}, K., \&
  {Brown}, W.~R. 2008{\natexlab{b}}, \aap, 488, L51

\bibitem[{{Sch{\"o}nrich} {et~al.}(2010){Sch{\"o}nrich}, {Binney}, \&
  {Dehnen}}]{schonrich10}
{Sch{\"o}nrich}, R., {Binney}, J., \& {Dehnen}, W. 2010, \mnras, 403, 1829

\bibitem[{{Sesana} {et~al.}(2009){Sesana}, {Madau}, \& {Haardt}}]{sesana09}
{Sesana}, A., {Madau}, P., \& {Haardt}, F. 2009, \mnras, 392, L31

\bibitem[{{Sills} {et~al.}(2009){Sills}, {Karakas}, \& {Lattanzio}}]{sills09}
{Sills}, A., {Karakas}, A., \& {Lattanzio}, J. 2009, \apj, 692, 1411

\bibitem[{{Tillich} {et~al.}(2009){Tillich}, {Przybilla}, {Scholz}, \&
  {Heber}}]{tillich09}
{Tillich}, A., {Przybilla}, N., {Scholz}, R., \& {Heber}, U. 2009, \aap, 507,
  L37

\bibitem[{{van der Marel}(2001)}]{vandermarel01}
{van der Marel}, R.~P. 2001, \aj, 122, 1827

\bibitem[{{Wilhelm} {et~al.}(1999)}]{wilhelm99b}
{Wilhelm}, R. {et~al.} 1999, \aj, 117, 2329

\bibitem[{{Yu} \& {Tremaine}(2003)}]{yu03}
{Yu}, Q. \& {Tremaine}, S. 2003, \apj, 599, 1129

\end{thebibliography}


\end{document}